\documentclass[10pt,twocolumn,showpacs,amsmath,amssymb,floatfix,superscriptaddress]{revtex4-2}

\usepackage{graphicx}
\usepackage{float}
\usepackage{dcolumn}
\usepackage{bm}
\usepackage{color} 
\usepackage{txfonts}
\usepackage{microtype}
\usepackage{hyperref}
\usepackage[english]{babel}
\usepackage{slashed}
\usepackage{gensymb}
\usepackage{epsfig}
\usepackage[normalem]{ulem}
\usepackage{array}
\usepackage{booktabs}
\usepackage[english]{babel}
\usepackage{multirow}
\allowdisplaybreaks[4]

\graphicspath{{/Users/zq671/Documents/BaiduSyncdisk/文章草稿/FireBall/manuscript/figs/}}	
\begin{document}
\renewcommand{\figurename}{FIG}	
\title{Generation of Polarized Overdense Pair-photon Fireball via Laser-Driven Nonlinear-linear QED Cascade}

\author{Zhen-Ke Dou\textsuperscript{\S}}
\affiliation{Ministry of Education Key Laboratory for Nonequilibrium Synthesis and Modulation of Condensed Matter, State key laboratory of electrical insulation and power equipment, Shaanxi Province Key Laboratory of Quantum Information and Quantum Optoelectronic Devices, School of Physics, Xi'an Jiaotong University, Xi'an 710049, China}
\altaffiliation{These authors contributed equally to this work.}
\author{Qian Zhao\textsuperscript{\S}}\email{zhaoq2019@xjtu.edu.cn}
\affiliation{Ministry of Education Key Laboratory for Nonequilibrium Synthesis and Modulation of Condensed Matter, State key laboratory of electrical insulation and power equipment, Shaanxi Province Key Laboratory of Quantum Information and Quantum Optoelectronic Devices, School of Physics, Xi'an Jiaotong University, Xi'an 710049, China}

\author{Feng Wan}
\affiliation{Ministry of Education Key Laboratory for Nonequilibrium Synthesis and Modulation of Condensed Matter, State key laboratory of electrical insulation and power equipment, Shaanxi Province Key Laboratory of Quantum Information and Quantum Optoelectronic Devices, School of Physics, Xi'an Jiaotong University, Xi'an 710049, China}

\author{Chong Lv}
\affiliation{Department of Nuclear Physics, China Institute of Atomic Energy, P. O. Box 275(7), Beijing 102413, China}

\author{Bing Guo}\email{guobing@ciae.ac.cn}
\affiliation{Department of Nuclear Physics, China Institute of Atomic Energy, P. O. Box 275(7), Beijing 102413, China}

\author{Jian-Xing Li}\email{jianxing@xjtu.edu.cn}
\affiliation{Ministry of Education Key Laboratory for Nonequilibrium Synthesis and Modulation of Condensed Matter, State key laboratory of electrical insulation and power equipment, Shaanxi Province Key Laboratory of Quantum Information and Quantum Optoelectronic Devices, School of Physics, Xi'an Jiaotong University, Xi'an 710049, China}	
\affiliation{Department of Nuclear Physics, China Institute of Atomic Energy, P. O. Box 275(7), Beijing 102413, China}
\begin{abstract}
Relativistic, polarized pair-photon fireballs are central to understand the microscopic energy transfer of high-energy astrophysical outflows, yet generating an overdense fireball in the laboratory, especially via an ultraintense laser, remains a formidable challenge. Here, we propose a novel method of laser-driven nonlinear-linear quantum electrodynamics (NL-QED) plasma, that dramatically lowers the laser intensity threshold for dense pair-photon fireball creation. By coupling polarization-resolved linear Breit-Wheeler and Compton processes with strong-field nonlinear radiation, we find that a self-organized NL-QED cascade is ignited in the laser-driven hole boring at intensities of $\sim 10^{22}~\mathrm{W/cm^2}$,  accessible with current 10-PW-class laser facilities. Consequently, we demonstrate the generation of a pair-photon fireball with an overdense gamma-ray bath (maximum average density $\overline{n_\gamma} \approx 3 \times 10^{22}~\mathrm{cm^{-3}}$) and a pair plasma reaching collective regime (maximum average density $\overline{n_\pm} \approx 3 \times 10^{17}~\mathrm{cm^{-3}}$), which is highly polarized.
Our method provides a comprehensive framework for studying laser-driven QED plasma and its application in laboratory astrophysics, probing multi-process QED physics.
\end{abstract}

\maketitle
Relativistic outflows or fireballs composed of electron-positron pairs and photons are central to high-energy astrophysics, providing the basic framework for interpreting gamma-ray bursts (GRBs), pulsar winds, active galactic nuclei, and compact-object mergers \cite{piran1999Gammaray,kumar2015physics,levinson2020Physics}. In these environments, enormous energy release in a compact region drives an initially optically thick pair-photon plasma whose subsequent expansion, acceleration, and radiative decoupling determine the observable high-energy signal \cite{uzdensky2014Plasma,anchordoqui2019Ultrahighenergy}. The key microphysics involves pair creation and annihilation, Compton scattering, bremsstrahlung, and radiation transport, which together regulate thermalization, opacity, and energy partition \cite{ruffini2010Electron}.  
In sufficiently dense flows, dissipation often proceeds through radiation-mediated shocks rather than collisionless shocks, while in more dilute or later stages electromagnetic instabilities and collisionless processes dominate \cite{lemoine2019Physics,levinson2020Physics,vanthieghem2022role,levinson2023Anomalous,faure2024Highenergy}. Interestingly, GRBs prompt and afterglow emissions show strong polarization, implying large-scale magnetic fields and anisotropic radiation in pair-photon outflows \cite{coburn2003Polarization,kalemci2007Search,wiersema2014Circular}. Beyond synchrotron emission, linear QED (L-QED) scattering of polarized soft photons by cold relativistic electrons, notably inverse Compton scattering, can reshape the high-energy polarization signal, providing a key probe of radiation transfer, magnetic geometry, and particle-photon coupling in pair-dominated jets \cite{chang2013Gamma,nava2016Linear}.
Understanding how a pair-rich, radiation-dominated plasma transitions between these regimes is essential for connecting the inner-engine physics to the nonthermal spectra, shock formation, and particle acceleration inferred from observations.

Ultraintense laser now offers a promising laboratory platform to investigate these pair-photon involved astrophysics \cite{dipiazza2012Extremely,mourou2019Nobel,zhang2020Relativistic,takabe2021Recent,sun2022Production,yu2024Bright}. Laser and beam driven interactions with high-$Z$ targets have demonstrated prolific positron production through Bethe-Heitler (BH) process \cite{chen2009Relativistic}, while recent advances have increased yields to the level required for collective pair-plasma behavior \cite{chen2015Scaling,sarri2015Generation,chen2023Perspectives,arrowsmith2024Laboratory}. These developments make it plausible to generate relativistic, quasi-neutral pair jets in the laboratory and to investigate their coupling to intense radiation fields, magnetic turbulence, and collisionless shock formation
 \cite{lobet2015Ultrafast,warwick2017Experimental}. Because of the limited pair yield per beam energy via BH process, producing the pair plasma with collective behavior is still a formidable challenge \cite{chen2023Perspectives,arrowsmith2024Laboratory}. 
Among the available routes, strong-field QED cascade are particularly promising \cite{nerush2011Laser,elkina2011QED,ridgers2012Dense,bulanov2013Electromagnetic,slade-lowther2019Identifying}.
Unlike the BH mechanism, which relies on material targets and therefore remains intrinsically baryon-coupled, strong-field QED cascade, consists of nonlinear Compton scattering (NCS) and nonlinear Breit-Wheeler (NBW), can develop in extreme-radiation plasma or near-vacuum environments \cite{qu2021Signature,mercuri-baron2025Growth,zaim2024Light}. They thus provide a natural pathway toward pair-photon plasmas in field-dominated environment, closer in spirit to magnetar magnetospheres and relativistic jet environments \cite{harding2006Physics,uzdensky2014Plasma,kumar2015physics}. 

However, self-sustained nonlinear QED avalanches generally require laser intensities approaching or exceeding \(10^{24}\,\mathrm{W/cm^2}\), which remain beyond current optical-laser capabilities \cite{yoon2021Realization,chen2023Perspectives}. This has motivated growing interest in lower-threshold channels for pair creation in laser-driven plasmas. In particular, recent studies have demonstrated that under accessible 10-PW conditions, the linear Breit--Wheeler (LBW) process can dominate pair production in laser--solid-density interactions \cite{he2021Dominance,song2024linear}. Furthermore, in near-critical-density plasma, LBW-dominated pair production was shown to be accompanied by positron acceleration to GeV energies \cite{sugimoto2023Positron}.  These results point to a distinct route to dense pair generation, driven by photon-photon collisions within the extreme-radiation plasma rather than by a purely nonlinear avalanche. Meanwhile, fully polarization-resolved treatments of linear Compton scattering (LCS) and LBW, together with their cascades, have now been established \cite{zhao2022Signatures,zhao2023Angledependent,zhao2023Cascade}. This makes it possible to incorporate polarized NL-QED cascade in laser-driven QED plasmas, and more broadly to explore the polarization-sensitive high-energy astrophysics, including polarized GRBs.

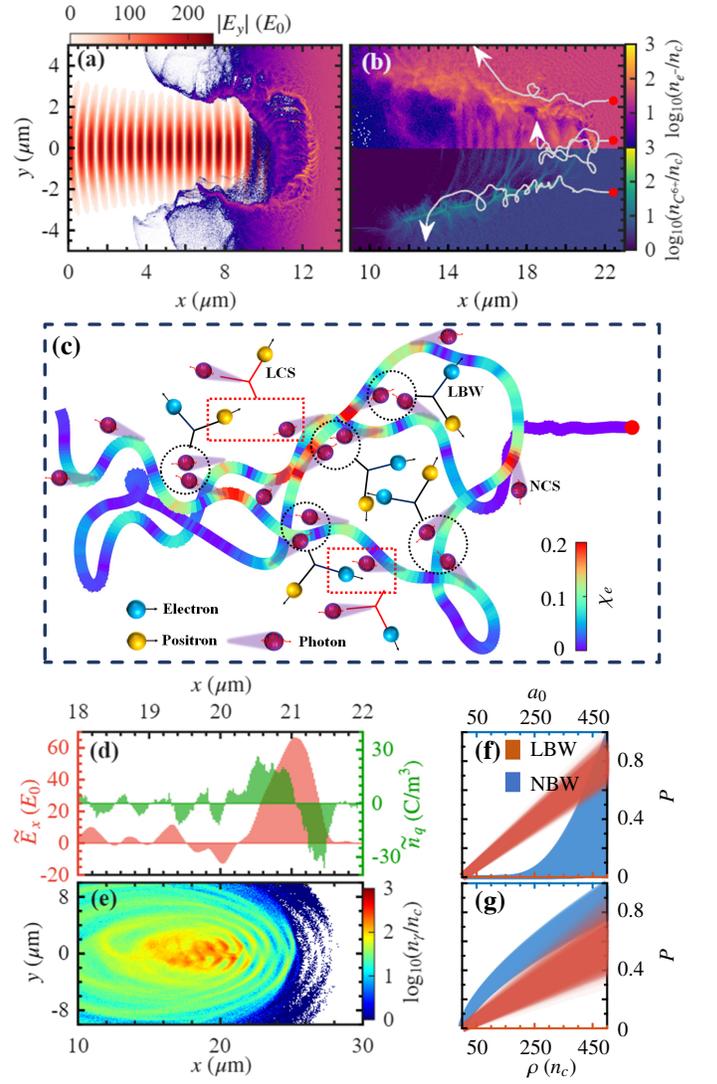
\begin{figure}[t!] 
\setlength{\abovecaptionskip}{-0.6cm}
	\centering
		\includegraphics[width=0.5\textwidth]{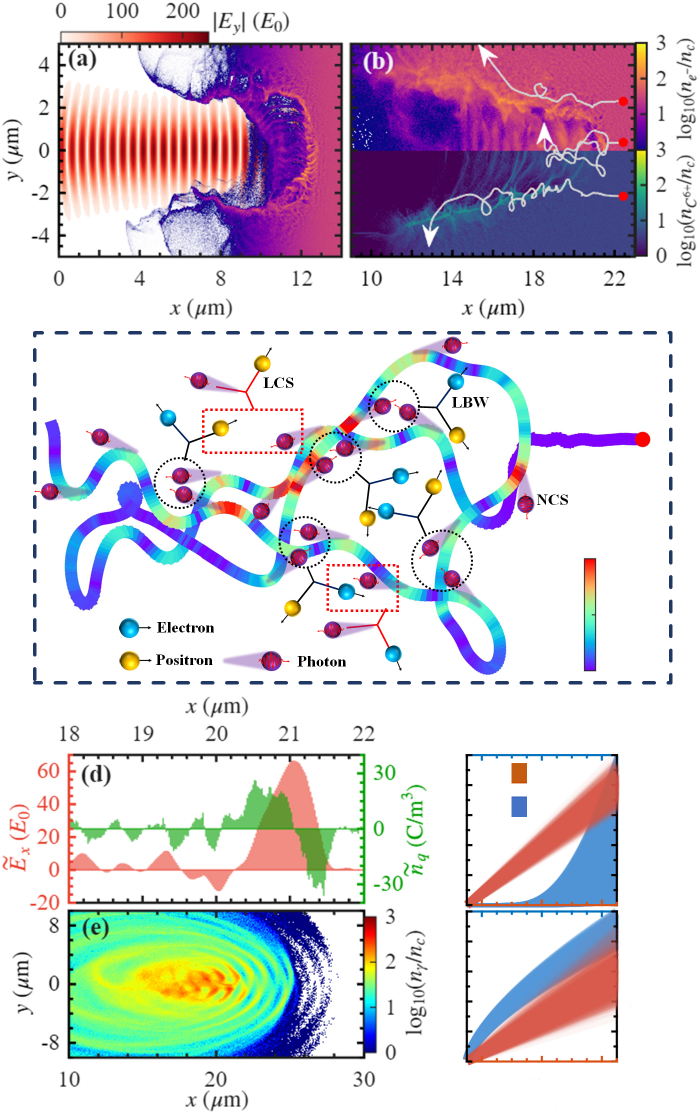}
\begin{picture}(300,25)
    \put(15,300){\normalsize\rotatebox{0}{\textbf{(c)}}}
    \put(175,147){\normalsize\textbf{(f)}}
    \put(175,90){\normalsize\textbf{(g)}}
    
    \put(222,205){\footnotesize\rotatebox{90}{$\chi_{e^-}$}}
    \put(205,186){\footnotesize{$0$}}
    \put(200,205){\footnotesize{$0.1$}}
    \put(200,225){\footnotesize{$0.2$}}
    
    \put(245,130){\footnotesize\rotatebox{90}{$P$}}
    \put(245,70){\footnotesize\rotatebox{90}{$P$}}
    \put(196,148){\footnotesize\text{LBW}}
    \put(196,135){\footnotesize\text{NBW}}
    \put(229,42){\footnotesize{$0$}}
    \put(229,65){\footnotesize{$0.4$}}
    \put(229,87){\footnotesize{$0.8$}}
    \put(229,100){\footnotesize{$0$}}
    \put(229,122){\footnotesize{$0.4$}}
    \put(229,144){\footnotesize{$0.8$}}
    \put(172,35){\footnotesize{$50$}}
    \put(192,35){\footnotesize{$250$}}
    \put(214,35){\footnotesize{$450$}}
    \put(195,28){\footnotesize{$\rho~(n_c)$}}
    \put(172,160){\footnotesize{$50$}}
    \put(192,160){\footnotesize{$250$}}
    \put(214,160){\footnotesize{$450$}}
    \put(195,170){\footnotesize{$a_0$}}
   \end{picture}
		\caption{Formation of the QED-active HB cavity and onset of the NL-QED cascade. (a) Laser field $|E_y|$ overlaid with electron density $n_{e^-}$ at $t=15~T_0$. (b) Electron density (upper panel) and carbon-ion density (lower panel) at $t=30~T_0$, together with three representative electron trajectories. (c) Enlarged view of the central trajectory in (b), color-coded by the electron quantum parameter $\chi_e$; instantaneous NCS and the associated LBW (dotted circular frame) and LCS (dotted rectangular frame) channels are indicated schematically. (d) Longitudinal electric field $\tilde{E}_x$ and charge density $\tilde{n}_q$, averaged over $y=0\pm1~\mu$m, at $t=30~T_0$. (e) Gamma-ray density (photons with energy $\varepsilon_\gamma>0.1$ MeV) at $t=30~T_0$. (f) Event probabilities $P$ of LBW and NBW versus background-photon density $\rho$ (bottom axis) and normalized laser amplitude $a_0$ (top axis), respectively, for projectile photon energies between 4 and 6 MeV; the background-photon energy in LBW varies from 1 to 500 MeV. (g) Same as (f), but for projectile-photon energies between 460 MeV and 560 MeV. The event probability $P$ is calculated by Eqs. (\ref{LBWprob}) and (\ref{NBWprob}).}
\label{fig:fig1}
\end{figure}

\begin{figure}[t!] 
	\centering
		\includegraphics[width=\linewidth]{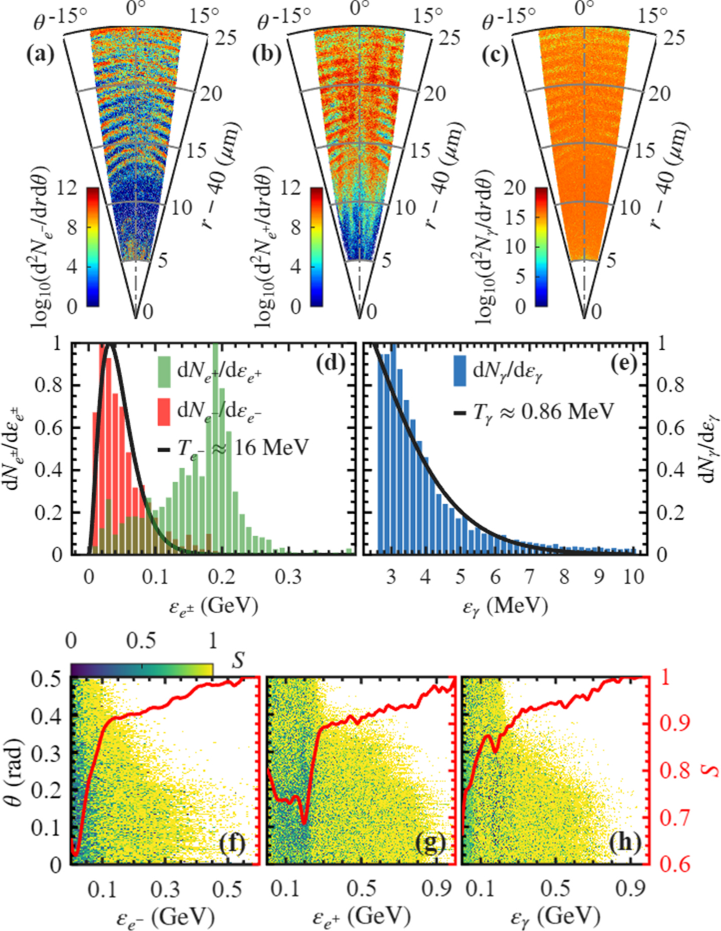}
		\caption{Spatial, spectral, and polarization structure of the final fireball at propagation time $t=70~T_0$. (a)-(c) Density distributions of electrons, positrons, and gamma photons in the $r$--$\theta$ plane of spherical coordinates. (d) Normalized energy spectra of electrons and positrons corresponding to (a) and (b), respectively; the electron spectrum is fitted by a Maxwell-J\"uttner distribution. (e) Normalized gamma-photon spectrum corresponding to (c), fitted by a Planckian distribution. (f)-(h) Polarization distributions of electrons, positrons, and gamma photons in phase space, respectively. Red-solid lines plot the energy-dependent polarization.}
\label{fig:fig2}
\end{figure}

\begin{figure}[t!] 
\centering
		\includegraphics[width=\linewidth]{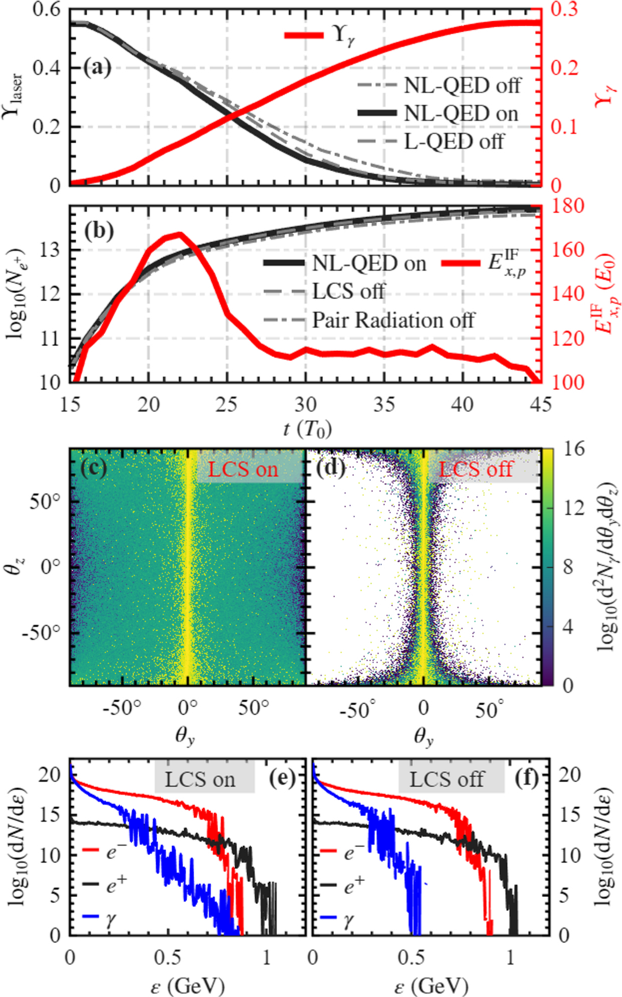}
		\caption{Energy conversion, pair production, and the role of L-QED processes. (a) Time evolution of the laser absorption rate $\Upsilon_{\rm laser}$ for three cases: NL-QED off, NL-QED on, and L-QED off, together with the gamma-photon energy fraction $\Upsilon_\gamma$. (b) Time evolution of the positron yield $\log_{10}(N_{e^+})$ for three cases: NL-QED on, LCS off, and pair radiation off, together with the peak interfacial field $E_{x,p}^{\rm IF}$. (c) and (d) Comparison of angular distributions of gamma-ray number, $\log_{10}[{\rm d}^2N_\gamma/({\rm d}\theta_y{\rm d}\theta_z)]$ at  intermediate time $t=30~T_0$, between the cases LCS on and LCS off, respectively. (e) and (f) Comparison of fireball spectra $\log_{10}({\rm d}N/{\rm d}\varepsilon)$  at intermediate time $t=30~T_0$, between the cases LCS on and LCS off, respectively.}
\label{fig:fig3}
\end{figure}

\begin{figure}[t!] 
	\begin{center}
		\includegraphics[width=\linewidth]{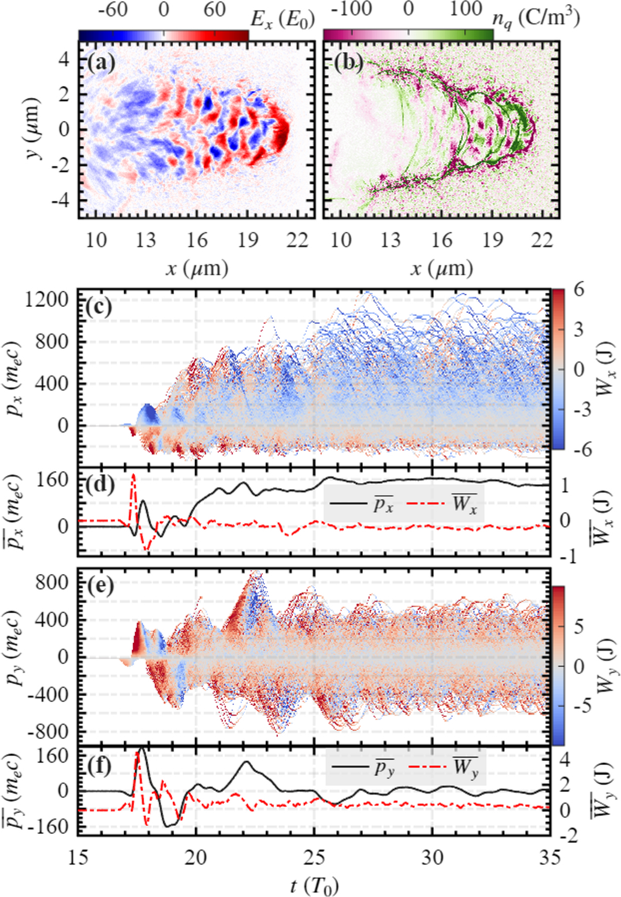}
		\caption{Recirculating acceleration inside the hole-boring cavity. (a) and (b) Snapshots of the longitudinal electric field $E_x$ and charge density $n_{q}$ at $t=30~T_0$. (c) Time evolution of the longitudinal momentum $p_x$ of sampled electrons, color-coded by the longitudinal work $W_x$; (d) corresponding ensemble-averaged longitudinal momentum $\overline{p_x}$ and work $\overline{W_x}$. (e) Same as (c), but for the transverse momentum $p_y$ and transverse work $W_y$; (f) corresponding ensemble-averaged transverse momentum $\overline{p_y}$ and work $\overline{W_y}$.}
\label{fig:fig4}
	\end{center}
\end{figure}

In this Letter, we propose a laboratory-accessible method of laser-driven QED plasma to generate a highly polarized pair-photon fireball with overdense gamma photons and quasi-neutral pair plasma, using currently operating 10-PW lasers, such as SULF and HPLS \cite{yamanouchi2021Progressa,radier202210}. By synergistically incorporating nonlinear and linear QED processes, our method dramatically lowers the intensity threshold normally required for pure strong-field QED cascade. We find that during the laser-driven hole boring (HB), disordered micro-sheath fields induce a stochastic recirculating heating of electrons [see Figs. \ref{fig:fig1}(a)-(d) and \ref{fig:fig4}].  This essential self-heating mechanism efficiently channels $\sim30\%$ of the laser energy into an overdense gamma-ray bath via extreme nonlinear radiation [see Fig. \ref{fig:fig1}(e) and Figs. \ref{fig:fig3}(a)-(b)], which subsequently triggers copious LBW pair creation and $e^\pm-$photon LCS, composing the NL-QED cascade. Consequently, the anisotropic gamma-ray bath together with its driven $e^+e^-$ pairs compose a quasi-spherical fireball with preserved polarization (see Fig. \ref{fig:fig2}). By switching off specific L-QED process in simulations, the essential roles of L-QED in the thermalization, angular redistribution, and polarization transfer of the fireball constituents are revealed [see Figs. \ref{fig:fig3}(c)-(f)]. The generated fireball demonstrates a laser-driven route to producing a pair plasma reaching collective regime \cite{supplement} and enables studies of pair-photon-driven plasma instabilities.

{\it Mechanism of fireball formation}.---Based on our developed QED-PIC code with polarization-resolved strong-field QED \cite{wan2023Simulations}, here we further incorporated L-QED into this code through binary-collision algorithm to consider the polarization-angle dependent LCS and LBW processes \cite{zhao2023Cascade}. Code benchmarks of production yield, energy-momentum conservation, and polarization distribution between theory and simulation by PIC are shown in Appendix \ref{yield}. The two-dimensional(2D) simulations use a moving window propagating along $+x$, with initial domain $0\le x\le25~\mu\mathrm{m}$ and $|y|\le10~\mu\mathrm{m}$, resolved by $750\times300$ cells. The target is a fully ionized hydrocarbon plasma with $n_{H^+}=n_{C^{6+}}=n_{e^-}/7$. Its density consists of a plateau at $n_0=30~n_c$ bounded by exponential ramps centered at $x_1=14~\mu\mathrm{m}$ and $x_2=34~\mu\mathrm{m}$ with $\sigma_x=3~\mu\mathrm{m}$. Where $n_c$ is critical density of laser with  $\lambda=1~\mathrm{\mu m}$ wavelength. The drive pulse is a tightly focused $p$-polarized Gaussian laser, 
 with spot size $w_0=1.5~\mu\mathrm{m}$, and peak amplitude $a_0=200$, corresponding to $I=5.52\times10^{22}~\mathrm{W/cm^2}$. Its temporal profile is $0.5[\tanh(2(t-1.5T_0))-\tanh(2(t-15.5T_0))]$, giving an effective duration of $14~T_0$, with $T_0=\lambda/c$ and $c$ the light speed in vacuum. These parameters result in a laser with peak power 1.9 PW and energy 87 J. Fields are normalized by $E_0=m_ec\omega/e$, where $m_e$ is electron mass, $e$ elementary charge, and $\omega=2\pi c/\lambda$ laser angular frequency. Electron quantum parameter is thus expressed as $\chi_{e^-} \simeq \frac{E \gamma_e (1 + \beta_e) \hbar\omega}{E_0m_ec^2}$ with electron Lorentz factor $\gamma _e$ and velocity $\beta_e$, and its experienced field $E$. A three-dimensional (3D) PIC simulation---which shows the consistent physical processes observed in 2D---is presented in the Supplemental Material \cite{supplement} as an independent validation.

At this intensity, the laser penetrates the overdense target through relativistic transparency and forms a strongly compressed electron sheath at the HB interface [Fig.~\ref{fig:fig1}(a)]. The sheath thickness is set by the relativistic skin depth, \(\ell_s^R\approx \sqrt{\gamma_0}\,c/\omega_p \simeq0.34~\mu\mathrm{m},\) where $\omega_p$ is the plasma frequency, and $\gamma_0=\sqrt{1+a_0^2/2}$ results from the ponderomotive approximation of hot electrons. Force balance between the ponderomotive drive and charge separation produces a giant longitudinal electrostatic field \(E_x^{\rm{IF}}\) at the HB interface. Estimating the ponderomotive potential as $\Phi_p\sim m_ec^2(\gamma_0-1)$ gives averaged \(\tilde{E}_x^{\rm{IF}}\approx \Phi_p/\ell_s^R \simeq 66~E_0,\) which is consistent with the simulated results in Fig. \ref{fig:fig1}(d). This field propels the carbon ions forward with feather-like density distribution, and acts as an electrostatic piston, intermittently ejecting downstream electrons into the ion cavity with density bunching current [Fig.~\ref{fig:fig1}(b)]. The bunched electrons distribute against the inhomogeneous ion background and generate internal micro-sheath fields, leading to the disordered longitudinal electric field $\tilde{E}_x$ [Fig.~\ref{fig:fig1}(d) and see 2D distribution in Fig.~\ref{fig:fig4}(a) below].  Note that the hydrogen ions are almost completely evacuated from the ion cavity during the early stage of HB.

Once injected, these electrons go through the recirculating acceleration inside the cavity, repeatedly interacting with the laser and the self-generated micro-sheath fields [Fig.~\ref{fig:fig1}(c)]. Their quantum parameter $\chi_{e^-}$ intermittently reaches values $\gtrsim0.1$, leading to extreme nonlinear radiation in anisotropic directions, creating gamma-ray bath with density over $10^3~n_c$ inside the cavity [Fig.~\ref{fig:fig1}(e)].  Such an extreme radiation originates from that $\tilde{E}_x$ repeatedly rephases bunched electrons relative to the laser, turning direct laser acceleration (DLA) into a stochastic heating process and extending the radiative lifetime of energetic electrons. The recirculating-heating mechanism in details---distinct from the magnetically assisted DLA \cite{sugimoto2023Positron}---is illustrated in Fig.~\ref{fig:fig4} below. The photon distribution in Fig.~\ref{fig:fig1} (e) reproduce the 3D simulation result with high fidelity, confirming that the recirculating-heating mechanism is inherently captured in 2D geometry \cite{supplement}.

The resulting overdense gamma-ray bath triggers copious collisions of LCS and LBW pair production [Fig.~\ref{fig:fig1}(c)]. According to the spectra [see Fig. \ref{fig:fig3}(e) below] of gamma-ray photons in Fig. \ref{fig:fig1}(e), theoretical estimate indicates that LBW dominates the pair production since the LBW probability can significantly exceed the NBW probability once normalized laser amplitude $a_0<300$ for a few of MeV projectile photons [Fig.~\ref{fig:fig1}(f)]. Even for hundreds of MeV projectile photons, LBW  probability can be comparable with NBW probability [Fig.~\ref{fig:fig1}(g)]. Detailed calculation of LBW and NBW probabilities is shown in Appendix \ref{probability}. Actually, the theoretical estimate is consistent with simulated results that there is negligible NBW pair production for the performed parameters Moreover, bremsstrahlung photons above $10~\mathrm{eV}$ are five orders of magnitude less dense than the NCS gamma-ray bath, and BH-produced positrons are about three orders of magnitude less dense than LBW-produced positrons \cite{supplement}; both processes are therefore negligible in our regime.

The gamma-ray bath propagates outside the subluminal HB cavity, accompanying the LBW pair productions of cross-collision photons  and $e^\pm$-photon LCS, forming a compact spherical pair-photon fireball [Figs.~\ref{fig:fig2}(a)-(c)]. The electron and positron densities retain radial modulation inherited from laser-cycle-scale acceleration, whereas the photon distribution is smoother because of multiple Compton scattering. Thermal expansion of the fireball--peak luminosity $\sim 2\times10^{21}~\mathrm{erg/s}$ at $t \simeq 40\,T_0$, with an overdense photon bath ($\overline{n_\gamma} \approx 30\,n_c$) and pair plasma ($\overline{n_\pm} \approx 3 \times 10^{-4}\,n_c$)—is tracked via its volume-averaged densities (Fig. S6 in \cite{supplement}).
The electron spectrum is well fitted by a Maxwell-J\"uttner distribution with $T_{e^-}\approx16~\mathrm{MeV}$, indicating efficient stochastic heating and Comptonization. The positron spectrum remains harder and less equilibrated, reflecting the propulsion of sheath fields [Fig.~\ref{fig:fig2}(d)]. The photon spectrum is nearly Planckian with $T_\gamma\approx0.86~\mathrm{MeV}$ over its bulk part [Fig.~\ref{fig:fig2}(e)], continuously replenished by the most energetic recirculating particles. The fireball remains highly polarized [Figs.~\ref{fig:fig2}(f)-(h)]: photon polarization grows with energy, reflecting the NCS polarization properties, while pair polarization transitions from NCS-dominated at high energies to LBW-mediated polarization transfer at low energies, with the electron--positron spectral discrepancy arising from electron thermalization and positron sheath-field acceleration. Multiple scattering broadens the angular distribution without erasing the polarization. The total averaged polarization of electrons, positrons, and photons are approximately 0.93, 0.91, and 0.93, respectively.

{\it Energy partition and role of linear QED channels}.---To quantify the global dynamics, we define the laser energy fraction as $\Upsilon_{\rm laser}=\mathcal{E}_L/\mathcal{E}_{0}$ and the gamma-ray energy fraction as $\Upsilon_\gamma=\mathcal{E}_\gamma/\mathcal{E}_{0}$, where $\mathcal{E}_{0}$ is initial total laser energy, and $\mathcal{E}_L$ and $\mathcal{E}_\gamma$ are the instantaneous total laser and photon energies. Once HB is established at $t\approx15~T_0$, the laser depletion rate increases sharply and the gamma energy grows nearly linearly until the pulse is exhausted at $t\approx40~T_0$ [Fig.~\ref{fig:fig3}(a)]. Relative to the case switching off NL-QED, the stronger depletion demonstrates that QED scattering opens an efficient radiative energy-loss channel. By contrast, switching off L-QED processes produces only a small change in the total absorption, indicating that the initial energy extraction from the laser is governed mainly by electron dynamics and nonlinear radiation. 

The pair yield tracks the temp-evolution amplitude of $E_{x}^{\rm IF}$, which is characterized by its peak $E_{x,p}^{\rm IF}$ [Fig.~\ref{fig:fig3}(b)]. The rise of $E_{x,p}^{\rm IF}$ is driven by laser self-focusing and sheath-front compression, while its decay coincides with the gradual saturation of pair production. $E_{x}^{\rm IF}$ therefore acts as the energy reservoir that regulates electron injection and radiative activity. Although the total positron yield is only weakly affected by switching off LCS and pair radiation, the photon phase space changes substantially, LCS broadens the gamma-ray angular distribution, enhances isotropization [Figs.~\ref{fig:fig3}(c)-(d)], and reshapes the spectrum through multiple LCS [Figs.~\ref{fig:fig3}(e)-(f)]. L-QED processes therefore play a limited role in primary energy extraction, but a major role in setting the thermodynamic and angular structure of the final fireball.

{\it Recirculating acceleration}.---The kinetic mechanism underlying the NL-QED cascade is disorder-assisted recirculating acceleration inside the HB cavity. When the HB develops at time about $t=10~T_0$, bunched electrons are injected into a highly nonuniform ion background, generating disordered longitudinal micro-sheath fields $E_x$ due to the layered charge density [Figs.~\ref{fig:fig4}(a) -(b)]. Rather than undergoing single-pass DLA, electrons are repeatedly reflected and reinjected by $E_x$. Meanwhile, the laser $\bm{J}\times\bm{B}$ force drives a longitudinal drift, creating a recirculating phase space that substantially prolongs electron residence time and sustains intense nonlinear radiation.

Particle tracking confirms this stochastic energization. The longitudinal work $W_x$ alternates in sign and averages to zero, indicating $E_x$ primarily recycles particles while net drift stems from magnetic-assisted laser acceleration [Figs.~\ref{fig:fig4}(c)-(d)]. Transversely, the laser-dominated work fluctuates heavily with vanishing ensemble averages [Figs.~\ref{fig:fig4}(e) and (f)]. Essentially, the cavity acts as a disorder-assisted laser heater: micro-sheaths continuously reset electron phases for repeated laser absorption, naturally explaining the broad electron spectra, the massive gamma-ray yield, and the consequent NL-QED cascade.

Featuring an ultraluminous and dense composition, the generated relativistic fireball provides a unique laboratory platform to investigate pair-radiation-driven current instabilities \cite{lemoine2019Physics,delgaudio2020Plasma}. To gauge the collective nature of the fireball pair plasma, we fit the co-moving momentum distribution to a relativistic Cauchy function \cite{arrowsmith2024Laboratory}. At $t = 50\,T_0$, with laser energy nearly depleted, the fit yields an effective temperature $\Theta \approx 4.5$ MeV and $\ell_D = \sqrt{\pi\Theta/4}\,\ell_s \approx 19~\mu\mathrm{m} < \ell_V = 20~\mu\mathrm{m}$, indicating approach of the collective condition. At $a_0 = 300$, the order-of-magnitude increase in pair density reduces $\ell_D$ to $\sim 9.4~\mu\mathrm{m}$, firmly establishing $\ell_D \ll \ell_V$. Compton cooling via LCS strips the high-energy spectral tail, driving the pairs toward isotropic equilibrium \cite{supplement}.

In summary, we have proposed a laboratory-accessible regime to generate an ultraluminous, highly polarized pair-photon fireball via a self-organized nonlinear-linear QED cascade. The underlying physics is governed by a disorder-assisted recirculating acceleration within a laser-driven hole boring, which efficiently channels $\sim30\%$ of the laser energy into an overdense gamma-ray bath, subsequently triggering massive linear-QED pair creation. This dynamics yields a compact, quasi-spherical fireball that intrinsically couples nonlinear radiation with linear scattering. Our findings establish a crucial experimental platform to probe strongly coupled QED plasmas, offering a microscopic window into the energy partitioning and polarization signatures of astrophysical outflows, such as gamma-ray bursts.

\vskip 0.5cm
{\it Acknowledgements}--- The work is supported by the National Natural Science Foundation of China (Grants No. 12425510, No. U2267204, No. 12441506, No. 12475249, No. 12447106, No. 12275209 , No. 12125509), the Science Challenge Project (No. TZ2025012), the National Key Research and Development (R\&D) Program (Grant No. 2024YFA1610900, No. 2024YFA1612700), the Innovative Scientific Program of CNNC, Natural Science Basic Research Program of Shaanxi (Grant No. 2024JC-YBQN-0042), and the Fundamental Research Funds for Central Universities (No. xzy012023046).

\appendix
\section{Benchmarks of NL-QED in PIC}\label{yield}

We benchmark the polarization-angle resolved Monte Carlo (MC) modules for LBW and LCS in the QED-PIC code \cite{wan2023Simulations}. In each cell, photons, electrons, and positrons are randomly paired using the no-time-count method. Each binary collision is then treated in the center-of-mass frame by MC sampling from completely polarized cross sections \cite{zhao2023Cascade}. Final-state four-momenta are Lorentz-transformed back to the laboratory frame. To handle unequal macro-particle weights while preserving event statistics, we adopt a multiplication-factor scheme that maintains energy-momentum conservation in each collision and improves sampling of secondary particles~\cite{Higginson2019}. After the L-QED sampling, newly produced and surviving particles are further checked for NCS or NBW events. Between time steps, particle trajectories follow the Lorentz equation and spin precession follows the Thomas-Bargmann-Michel-Telegdi equation.

The L-QED modules are validated against theory in three aspects: production yield, energy--momentum conservation, and polarization transfer. The event rate is $\frac{dR}{dt}=L\sigma$, where $L$ is the luminosity and $\sigma$ is the corresponding cross section~\cite{Herr2003Concept}. In the benchmark runs, we simulate photon--photon, photon--electron, and photon--positron beam collisions using the three-dimensional SLIPs code. The box size is $6\times4\times4~\mu{\rm m}^{3}$, resolved by $120\times80\times80$ cells, with 10 macro-particles per cell. For statistical convergence, each cylindrical beam has radius $r=1~\mu{\rm m}$, length $l=1~\mu{\rm m}$, uniform density $1000~n_c$, and luminosity $L=\frac{N_1N_2f}{\pi r^2}$. The cross sections used in the theoretical calculations below are detailed in \cite{zhao2023Cascade}.

\begin{figure}[t!]
    \centering
    \includegraphics[width=0.95\linewidth]{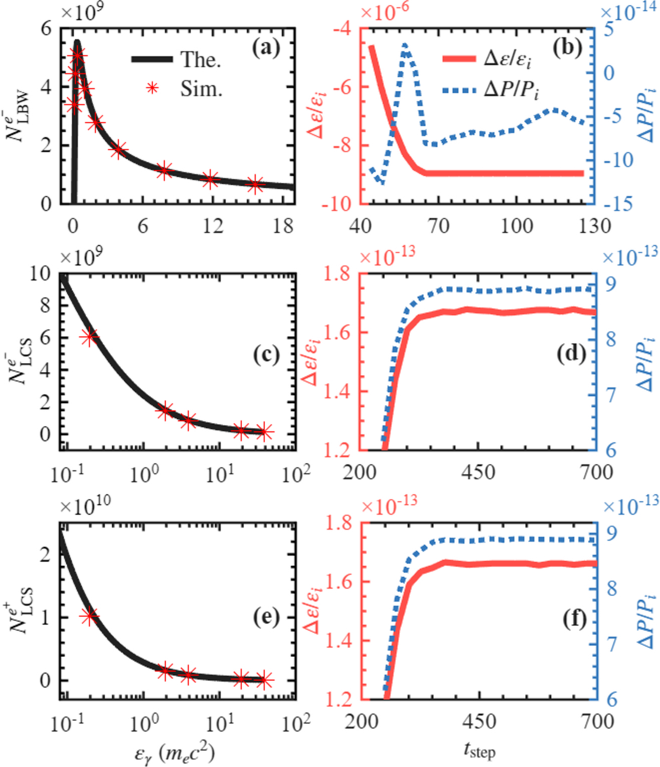}
    \caption{Benchmark of cross sections and conservation laws. (a), (c) and (e) Theoretical curves (The.) and simulated results (Sim.) for the yields of LBW electron,  LCS electron, and LCS positron versus incident gamma energy. (b), (d), (f) Relative variations of the total energy and momentum varying with simulating time steps, during beam-beam collisions corresponding to (a), (c) and (e), respectively.}
    \label{figS1}
\end{figure}
In order to examine simulated yields of LCS and LBW with theoretical calculation, the colliding beams are initialized as a monoenergetic target beam with fixed energy, a monoenergetic projectile beam with different  energy points. The simulated yields agree well with the theoretical expectations over the relevant energy range, confirming the implemented LBW and LCS cross sections [Figs. \ref{figS1}(a),(c) and (e)]. Additional tests with two monoenergetic beams of 2 and 4 MeV show negligible variations of the total energy and momentum over many computational steps, demonstrating accurate conservation [Figs. \ref{figS1}(b),(d) and (f)]. 

\begin{figure}[t!]
    \centering
    \includegraphics[width=0.95\linewidth]{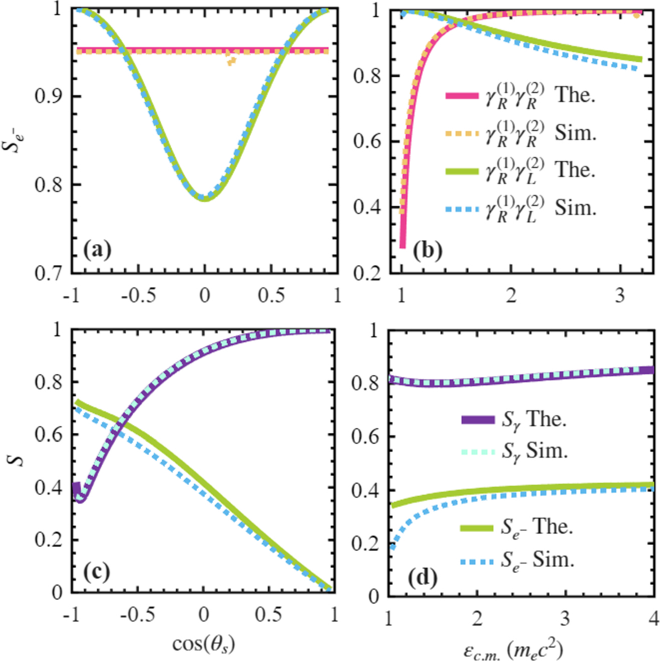}
    \caption{Benchmark of polarization transfer. Comparison between theory and PIC simulation for the polarization of final electrons or photons as functions of scattering angle and center-of-momentum energy: (a) and (b) Polarization of LBW electron for collisional scenarios of $\gamma_R^{(1)}\gamma_R^{(2)}$ and $\gamma_R^{(1)}\gamma_L^{(2)}$; (c) Angle-dependence of polarization for final-state photon $S_{\gamma}$ and electron $S_{e^-}$ from LCS. (d) Similar to (c) but for energy dependence. Polarization of LCS is calculated with initial polarization of photons $s=(-0.43,-0.75,0.5)$, and unpolarized electrons.}
    \label{figS2}
\end{figure}

For examination of polarization transfer in LCS and LBW, we set up two colliding beams, with uniform energy distribution between 0.1 MeV and 2 MeV. By denoting $\gamma_R^{(1)}$ as one photon beam with right-hand circular polarization and $\gamma_L^{(2)}$ as another photon beam with left-hand circular polarization, two colliding scenarios for LBW are examined, namely $\gamma_R^{(1)}\gamma_R^{(2)}$ and $\gamma_R^{(1)}\gamma_L^{(2)}$.
See that both the energy dependence and angle dependence of pair polarization $S_{e^-}$ from simulations agree well with the theoretical calculations [Figs. \ref{figS2}(a)-(b)]. For examination of LCS, we consider collision between photon beam with arbitrary polarization and unpolarized electron beam. Both the angle-dependence and energy-dependence of final-state polarization, from the MC simulation reproduce the analytical polarization distributions [Figs. \ref{figS2}(c)-(d)]. The slight discrepancy is mainly due to the limited numbers of produced particles in the simulation, since the theoretical polarization is the statistical average from simulated particles.
\section{Probabilities of LBW and NBW}\label{probability}

To estimate the probability of LBW pair creation in the simulation, we use the positron yield from a single photon-beam collision~\cite{esnault2021Electronpositron},
\begin{equation}
N^{e^+}=L_{12}\sigma_{\gamma\gamma}^{\rm int},
\end{equation}
with geometric luminosity
\begin{equation}
L_{12}=c(1-\cos\psi_{12})N_1N_2\!\int\!\rho_1\rho_2\,d^3Vdt,
\end{equation}
and energy-integrated cross section
\begin{equation}
\sigma_{\gamma\gamma}^{\rm int}=\iint f_1(\varepsilon_1)f_2(\varepsilon_2)\sigma_{\gamma\gamma}\,d\varepsilon_1d\varepsilon_2.
\end{equation}
Here $\psi_{12}$ is the collision angle, $N_i$, $\rho_i$, and $f_i$ are the photon number, density, and spectrum of beam $i$, respectively. The LBW cross section in the center-of-mass frame is
\begin{eqnarray}\label{LBWprob}
\sigma_{\gamma\gamma}&=&4\pi r_e^2\frac{m_e^2c^4}{s}
\left[
\left(2+\frac{8m_e^2c^4}{s}-\frac{16m_e^4c^8}{s^2}\right)
\ln\frac{\sqrt{s}+\sqrt{s-4m_e^2c^4}}{2m_ec^2}\right.\nonumber\\
&-&\left.\sqrt{1-\frac{4m_e^2c^4}{s}}
\left(1+\frac{4m_e^2c^4}{s}\right)
\right],
\end{eqnarray}
where $s=2\varepsilon_1\varepsilon_2(1-\cos\psi_{12})=\varepsilon_{c.m.}^2$.

In practice, the photon density is taken as uniform in each cell. For Fig.~\ref{fig:fig1}(f), we evaluate collisions between photons in the 4--6 MeV range and photons in the 1--500 MeV range; for Fig.~\ref{fig:fig1}(g), the high-energy interval is 460--560 MeV. For each case, we compute $\sigma_{\gamma\gamma}$ for all relevant energy pairs, integrate over the spectra, evaluate the cell luminosity from the local density $\rho$, and randomly sample $\psi_{12}\in[0,\pi]$. The resulting positron yield is normalized to its maximum value and used as the relative LBW probability.

The NBW probability is evaluated from the standard differential rate~\cite{wan2023Simulations},
\begin{equation}\label{NBWprob}
    \frac{d^2W}{d\varepsilon_+dt}=16W_p\Big[{\rm IntK_{\frac{1}{3}}(\zeta)}+\frac{\varepsilon^2_++\varepsilon^2_-}{\varepsilon_+\varepsilon_-}{\rm K_{\frac{2}{3}}(\zeta)}\Big]
\end{equation}
where $W_p=\alpha m_e^2c^4/(16\sqrt{3}\pi\hbar\varepsilon_\gamma^2)$, $\varepsilon_\gamma=\varepsilon_++\varepsilon_-$, and $\zeta=2\varepsilon_\gamma^2/(3\chi_\gamma\varepsilon_+\varepsilon_-)$. Using this rate, we calculate the positron-production probability for photons in the 4--6 MeV and 460--560 MeV ranges under different field intensity $a_0$, as shown in Figs.~\ref{fig:fig1}(f) and (g).

\bibliography{library}

\end{document}